\documentclass[12pt,preprint]{aastex}
\usepackage{emulateapj5}

\newcommand{\der}{{\rm d}}
\newcommand{\Mc}{M_{\rm i}}

\newcommand\tc{t_{\rm i}}
\newcommand{\Mi}{M_{\rm f}}
\newcommand{\Vc}{V_{\rm c}}
\newcommand{\Ri}{R_{\rm f}}
\newcommand{\ti}{t_{\rm f}}
\newcommand{\ri}{r^{\rm i}}
\newcommand{\rs}{r_{\rm s}}
\newcommand{\rf}{r^{\rm f}}
\newcommand{\rc}{r^{\rm c}}
\newcommand{\ra}{r^{\rm a}}
\newcommand{\fric}{_{\rm fric}}
\newcommand{\orb}{_{\rm dec}}
\newcommand{\last}{_{\rm last}}
\newcommand{\merg}{}

\newcommand{\rel}{_{\rm rel}}

\newcommand{\minb}{_{\rm min}}
\newcommand{\sat}{_{\rm s}}
\newcommand{\roc}{\rho_{\rm c}}
\newcommand{\rou}{\rho_{\rm u}}
\newcommand{\rocr}{\rho_{\rm crit}}
\newcommand{\delc}{\delta_{\rm c}}
\newcommand{\Delc}{\Delta}
\newcommand{\delm}{\Delta_{\rm m}}
\newcommand{\PS}{_{\rm LC}}
\newcommand{\nbody}{{$N$}-body }

\shorttitle{Halo Structure}
\shortauthors{Manrique et al.}

\begin{document}


\title{On the Origin of the Inner Structure of Halos}


\author{Alberto Manrique, Andreu Raig, Eduard
 Salvador-Sol\'e\altaffilmark{1}, Teresa Sanchis \\
and Jos\'e Mar\'\i a Solanes} 
\affil{Departament d'Astronomia i Meteorologia, Universitat de Barcelona, 
\\Mart{\'\i} Franqu\`es 1, E-08028 Barcelona, Spain}  
\email{Alberto.Manrique@am.ub.es, araig@am.ub.es, eduard@am.ub.es, 
tsanchis@am.ub.es, and jsolanes@am.ub.es}


\altaffiltext{1}{CER for Astrophysics, Particle Physics, and
Cosmology, Universitat de Barcelona, Mart{\'\i} Franqu\`es 1, E-08028
Barcelona, Spain}

\begin{abstract}

We calculate by means of the Press-Schechter formalism the density
profile developed by dark-matter halos during accretion, i.e., the
continuous aggregation of small clumps. We find that the shape of the
predicted profile is similar to that shown by halos in high-resolution
cosmological simulations. Furthermore, the mass-concentration relation
is correctly reproduced at any redshift in all the hierarchical
cosmologies analyzed, except for very large halo masses. The role of
major mergers, which can cause the rearrangement of the halo structure
through violent relaxation, is also investigated. We show that, as a
result of the boundary conditions imposed by the matter continuously
infalling into the halo during the violent relaxation process, the
shape of the density profile emerging from major mergers is essentially
identical to the shape the halo would have developed through pure
accretion. This result explains why, according to high-resolution
cosmological simulations, relaxed halos of a given mass have the same
density profile regardless of whether they have had a recent
merger or not, and why both spherical infall and hierarchical assembly lead to
very similar density profiles. Finally, we demonstrate that the density
profile of relaxed halos is not affected either by the capture of
clumps of intermediate mass.

\end{abstract}




\section{INTRODUCTION}\label{intro}

The dominant dark component of matter in the universe is clustered in
bound halos which form the skeleton of all astronomical objects of
cosmological interest, from dwarf galaxies to rich galaxy clusters.
The determination of the inner structure of such halos has been
addressed both analytically and by numerical simulations.

The analytical approach to this problem was pioneered by Gunn \& Gott
(1972), who adopted the simplifying assumption that halos grow through
spherical infall, i.e., the monolithic collapse of a density
fluctuation of isotropically distributed, cold, collisionless matter
in an otherwise homogeneous expanding universe. This allowed them to
derive the density profile resulting from self-similar initial
conditions in an Einstein-de Sitter universe up to the onset of shell
crossing. The effects of shell crossing and of adopting more and more
realistic initial conditions were addressed and the whole treatment
was refined in a series of subsequent papers
\citep{gott75,gunn77,fg84,berts85,hs85,rg87,ryd88,zh93,hw99,lok00,LH00,
pgrs,ekp,nuss}.

According to these studies, after the collapse of some initial seed, a
stationary regime is reached, the so-called secondary infall, in which
the system grows inside-out. This is the consequence of the fact that
the orbital period of particles in any shell is smaller than the
characteristic time of secular variation of their apapsis (or
turn-around radius) owing to shell crossing of the infalling layers.
Accordingly, 1) particles spend most of the time near the apapsis of
their orbits, which makes their time-averaged radius close to that
value, and 2) shell crossing gently alters the initial apapsis
(corresponding to maximum expansion) of layers, making them contract
by a monotonously varying factor of order two and, hence, with no
apapsis crossing. This is an important result since, provided the
infall rate of matter is known, the inside-out growth condition
completely determines the halo density profile. Much progress has been
achieved in the last twenty years in the modeling of halo mass
growth. In particular, the extended Press-Schechter (PS) model
\citep{PS,b,BCEK,LC93} makes accurate predictions on the rate at which
halos increase their mass in hierarchical cosmologies
\citep{LC94}. Unfortunately, in such cosmologies, halos develop from
small to large scales by successive aggregations rather than through
smooth spherical infall.

There are different ways in the literature to refer to the mass
assembly process. Some authors call each individual aggregation event
``merger'' (e.g., \citealt{b,BCEK,LC93}), while others use the word
``accretion'' to refer to their cumulative effect (e.g.,
\citealt{Wechs02}, hereafter W02; \citealt{zhao03}). In $N$-body or
Monte Carlo simulations, it is also usual to use the words ``merger''
and ``accretion'' depending on whether the captured clumps are resolved
or not, respectively (e.g., \citealt{som99}). On the other hand, if one
is only interested in the {\it mass growth} of halos, the distinction
between ``major'' and ``minor'' mergers or ``fast'' and ``slow''
accretion is not important as it represents solely an arbitrary
classification of captures according to the mass increase they
produce. However, in dealing with the structural and kinematical
effects of aggregation events on a given halo, such a distinction is
crucial.

When a halo captures a much less massive clump its structure is
essentially unperturbed while, in the case that both objects have
comparable masses, the system is brought far from equilibrium and
violently relaxes again \citep{LB67}, thus loosing the memory of the
previous history. The new equilibrium state is characterized by a
normal distribution of the velocities of the different constituents
similar to that yielded by two-body relaxation, although independent
of the particle mass. Consequently, were halos unperturbed after any
such dramatic event, they would end up as spherical systems with
uniform, isotropic velocity dispersion. For this reason, halos are
often modeled as isotropic, monomassic, isothermal spheres (e.g.,
\citealt{King72,SIR99}). However, halos are not isolated systems and
they continue collecting matter during violent relaxation, which makes
the final state difficult to predict (e.g., \citealt{s87}). In
addition, the frontier between minor and major captures is rather
blurred and intermediate captures have their own specific dynamical
effects. The associated clumps undergo substantial dynamical friction
and spiral to the halo center rather than move along stable
orbits. Then, the density distribution is neither unaltered nor
completely rearranged, but becomes a little cuspier at the halo
center.

The situation is therefore quite complicated and difficult to
implement in an analytical model. For this reason some authors have
turned to numerical experiments. Navarro, Frenk, \& White (1996; 1997,
hereafter NFW) found, through \nbody simulations of several
hierarchical cosmologies, that the spherically averaged density
profile of current halos of a wide range of masses is always well
fitted by the simple analytical expression
\begin{equation}
\rho(r)={\roc \rs^3 \over r(\rs + r)^2}\,, 
\label{nfw}
\end{equation} 
hereafter referred to as the NFW profile. In equation (\ref{nfw}), $r$
is the radial distance to the halo center, and $\rs$ and $\roc$ are
the halo scale radius and characteristic density, respectively. The
latter two parameters are related to each other and to the mass $M$ of
the halo through the condition that the average density within the
virial radius $R$ of the system is equal to $f$ times $\rou$, with $f$
a constant factor independent of mass and $\rou$ some characteristic
density of the universe. NFW adopted $\rou$ equal to the critical
density $\rocr$, and $f=200$, although $\rou$ equal to the mean cosmic
density $\bar\rho$ and other values of $f$ are also often adopted.

On the other hand, \citet{CL97} showed that the spherically averaged,
locally isotropized, velocity dispersion profile $\Sigma(r)$ is well
fitted by the solution of the Jeans equation for hydrostatic
equilibrium and negligible rotation
\begin{equation}
\Sigma^2(r)\left( {\der \ln \Sigma^2 \over \der \ln r} + 
{\der \ln \rho \over \der \ln r} \right) = 
- {3\,G M(r) \over r}\,,
\label{heq}
\end{equation}
where $G$ is the gravitational constant, $\rho(r)$ the density profile
given in equation (\ref{nfw}) and $M(r)$ the corresponding mass
profile, for the boundary condition of null pressure at infinity.

These results have been extensively confirmed (e.g.,
\citealt{TBW97,HJS99a,Bull01}). There is only some controversy on the
exact slope of the density profile at very small radii
\citep{Moore98,JS00,FM01,K01,P03}, the most affected by the limited
resolution of simulations. Nonetheless, in spite of the substantial
progress in our knowledge on the inner structure of halos afforded by
\nbody simulations, the physical origin of the halo density and
velocity dispersion profiles remains poorly understood.

Another well known result inferred from $N$-body simulations is that,
in all the cosmologies investigated, the halo concentration $c\equiv
R/\rs$ decreases with increasing halo mass. NFW proposed that this is
a consequence of the fact that, in hierarchical cosmologies, more
massive halos form later when the mean density of the universe is
lower. They demonstrated that the empirical mass-density relation is
automatically recovered when the proportionality $\roc\propto
\bar\rho(\ti)$ is assumed. They were forced, however, to define the
formation time $\ti$ as the time some progenitor collected {\it one
hundredth\/} of the final halo mass. Salvador-Sol\'e, Solanes, \&
Manrique (1998, hereafter SSM) showed, nonetheless, the robustness of
that result using a better-motivated definition of the halo formation
time: the last time the system was rearranged in a major merger. They
also showed that the relation $\roc\propto \bar\rho(\ti)$ (strictly
with $\rocr$ instead of $\bar\rho$, though this makes no difference
for the typical halo formation times in any popular cosmology) is
equivalent to a universal dimensionless density profile for newborn
halos, thus bringing some physical motivation to the assumed
proportionality. Unfortunately, the $z$-dependence of the
concentration predicted by such an assumption is not corroborated by
\nbody simulations (Bullock et al.~2001, hereafter B01).

A more promising result found by \citet{arfh} was that the density
profile of halos that have not been subjected to any major merger is
very similar to the NFW profile. As discussed below, the continuous
aggregation of small clumps proceeds essentially as secondary
infall. It is therefore well understood why, using the spherical infall
model from quite a realistic density profile of the initial
perturbation such as the typical density run around peaks, \citet{pgrs}
also found a density profile similar to the NFW one. On the other hand,
since halos grow inside-out during secondary infall, these results are
also consistent with the finding by \citet{ns99} and \citet{kull00}
that the density profile of halos growing inside-out (or through stable
clustering as the former authors call this process) and having not
experienced a major merger is similar to the NFW profile.

However, the density profiles obtained from spherical infall do not
satisfy the correct mass-concentration relation: the predicted
concentration is substantially higher than that of simulated
halos. Therefore, the role of spherical infall in the density profile
of halos is still debatable. In this one respect, is also important
mentioning that the inside-out growth condition is inconsistent with
the dynamical effects of intermediate mass captures as mid-size clumps
tend to migrate to the halo center. \citet{SW98} have examined the
possibility that the NFW profile results from the repeated action of
this kind of captures, concluding that it might explain the central
cusp (see also \citealt{SCO00}). But, as shown by \citet{ns99}, when
the specific effects of intermediate captures are taken into account,
the resulting density profile deviates more from the NFW profile than
the one obtained under strict accretion ---the role, in this scheme, of
the tidal stripping of captured clumps has been also examined by
\citet{DDH03,Detal03}. Thus, before spherical infall can be confirmed
as the main responsible for the density profile of halos, it is
necessary to understand why the alterations induced by intermediate
captures do not show up more frequently.

More importantly, all these works do not take into account the dramatic
effects of major mergers (or fast accretion). By definition, this
process causes the rearrangement of the system destroying in this way
the density profile developed until that moment by spherical infall. Of
course, this would go unnoticed if the density profile emerging from
violent relaxation turns out to be very similar to the one a halo of
the same mass would have developed through spherical infall.
Surprisingly enough, numerical simulations indicate that this is just
what happens. Indeed, high-resolution \nbody simulations show that
``the occurrence of a recent [major] merger is not an important factor
affecting the [shape and the] concentration'' of simulated halos (W02,
p. 568). Likewise, the phase space density, $\rho(r)/\Sigma^3(r)$, of
halos in simulations of hierarchical cosmologies is found to follow,
independently of their assembly history, a power-law for more than two
decades in radius with logarithmic slope close to that predicted by
\citet{berts85} using the spherical infall model (\citealt{TN01}).
Furthermore, the density profile of halos obtained in numerical
simulations of monolithic collapse are very similar to those obtained
in hierarchical cosmologies, which indicates that ``the merger history
does not play a role in determining the halo structure''
(\citealt{Moore99}, p. 1147; see also \citealt{HJS99b}). In other
words, the density profile of simulated halos is always the same
regardless of whether (and when) their structure has been rearranged by
major mergers. This necessarily implies that the density profile
emerging from violent relaxation must be very similar to that developed
through spherical infall. But what is the reason for such a
coincidence?

The present paper attempts to answer all these questions. We begin in
\S~\ref{mps} by describing the basic equations of a variant of the
extended PS model that are used in the computations of subsequent
sections. We then examine, in \S~\ref{acc}, whether pure accretion,
which yields density profiles of the NFW type, can recover the
mass-concentration relation of simulated halos. In \S~\ref{merg}, we
analyze the effects of mergers and intermediate captures. The results
are summarized and discussed in \S~\ref{diss}.

\section{MASS GROWTH THROUGH ACCRETION AND MERGERS}\label{mps}

The Modified Press-Schechter (MPS) model (SSM; \citealt{rgs98}, and
2001, hereafter RGS) is a variant of the extended PS model intended to
describe not only the mass growth of dark-matter halos in hierarchical
cosmologies, but their inner structure as well. To this end it
distinguishes between major and minor aggregation events (i.e., major
and minor mergers or fast and slow accretion) according to whether they
cause or not the complete rearrangement of the system, based on the usual
comparison of the resulting fractional mass increase with respect to
the reference halo with some pre-established threshold, $\delm$,
separating the two regimes.  Note that a given aggregation event may be
seen as major or minor depending on the partner halo considered. For
consistency, we say that halos are destroyed in major aggregation
events while they survive in minor ones. We will follow here the most
usual notation adopted in observational studies (and in SSM) and refer
to major aggregation events as ``mergers'' and to the continuous
aggregation of small clumps as ``accretion'' (one should avoid
confusion with the different meaning of these words in other works;
\S~\ref{intro}).

According to the definition of $\delm$, the accretion rate, that is,
the rate at which halos with $M$ at $t$ increase their mass between
two mergers, is
\begin{equation}
\ra(M,t)=\int_M^{M(1+\delm)}\,(M'-M)
\,r\PS(M, M',t)\,\der M'\, , \label{accr}
\end{equation}
with $r\PS(M, M',t)\;\der M'$ the instantaneous transition rate at $t$
from halos with $M$ to halos between $M'$ and $M'+\der M'$, provided
by the extended PS model \citep{LC93}
\begin{eqnarray}
\hskip-2pt r\PS(M, M',t)&=&
{\sqrt{2/\pi}
\over\sigma^2(M')}
{\der\delc\over \der t}{\der\sigma(M')\over\der M'} 
\left[1-{\sigma^2(M')\over\sigma^2(M)}\right]^{-3/2} \nonumber\\ 
&\times&\exp\left\{-{\delc^2(t)\over 2\sigma^2(M')}\left
[1-{\sigma^2(M')\over\sigma^2(M)}\right]\right\}\,.
\label{umr} 
\end{eqnarray}
In equation (\ref{umr}), $\delc(t)$ is the linear extrapolation to the
present time $t_0$ of the critical overdensity of primordial
fluctuations collapsing at $t$, and $\sigma(M)\equiv \sigma(M,t_0)$ is
the r.m.s. fluctuation of the density field at $t_0$ smoothed over
spheres of mass $M$. Both $\delc(t)$ and $\sigma(M)$ depend on the
cosmology. The $M(t)$ track followed, during accretion, by halos with
$\Mc$ at $\tc$ is therefore the solution of the differential equation
\begin{equation}
{\der M\over\der t}=\ra[M(t),t] \label{act}
\end{equation}
for the initial condition $M(\tc)=\Mc$. Strictly speaking, this is the
{\it average} track followed by those halos. Real accretion tracks
actually diffuse from it owing to the random effects of individual
captures. The scatter remains nonetheless quite limited along the
typical lifetime of halos (see RGS).

The destruction of a halo does not necessarily imply the formation of
a new one: the largest partner participating in the merger can
perceive it as minor and survive. Only those mergers in which {\it all\/}
initial halos are destroyed or, equivalently, none of them survives do
mark the formation of a new halo. (The definitions of halo destruction
and survival imply, indeed, that there is at most one surviving halo
in any aggregation event.) Thus, the formation of a halo corresponds
to the last time the system was rearranged in a merger. 

To derive the formation rate we need to introduce the capture rate,
$\rc(M',M,t)\,\der M$, giving the rate at which a halo with mass $M'$
at $t$ results from a major merger of a halo with mass $M$ to
$M+\der M$
\begin{eqnarray}
\rc(M',M,t)\,\der M\,\der t &=& r\PS(M,M',t)\,
\theta[M'-M(1+\delm)] \nonumber \\
&\times&\frac{N(M,t)}{N(M',t)}\,\der M\,\der t
\label{cap}\,,
\end{eqnarray}
with $r\PS(M,M',t)$ given by equation (\ref{umr}), $N(M,t)$ the PS
mass function, and $\theta(x)$ the Heaviside function. As shown by
RGS, mergers leading to the formation of new halos are essentially
binary. This is consistent with the fact that the capture rate is
closely symmetrical around $M=M'/2$ (see Fig.~\ref{1} in RGS), at
least in the range $[M'\delm/(1+\delm),M'/(1+\delm)]$ corresponding to
captures leading to the formation of new halos. Note that a fractional
mass relative to the initial object equal to $\delm$ represents a
fractional mass relative to the final object equal to
$\delm/(1+\delm)$. Since each symmetric pair of such captures produces
the formation of one new halo with mass $M'$, the formation rate of
halos of this mass is
\begin{equation} 
\rf(M',t)= {1\over 2}\, \int_{M'\delm/(1+ \delm)}^{M'/(1+\delm)} r\PS
(M,M',t)\,{N(M,t) \over N(M',t)}\,\der M\label{form}\,.
\end{equation}

The probability distribution function (PDF) of formation times can then be
derived by taking into account that the mass evolution of a halo since
its formation is given by the accretion track $M(t)$ solution of
equation (\ref{act}). Thus, the cumulative number density of halos at
$t_{\rm i}$ with masses in the arbitrarily small range $M_{\rm i}$ to
$M_{\rm i}+\delta M_{\rm i}$ that pre-exist at a time $t<t_{\rm i}$
or, equivalently, the spatial number density of halos that evolve by
accretion from $t$ to $t_{\rm i}$ ending up with a mass between
$M_{\rm i}$ and $M_{\rm i}+\delta M_{\rm i}$ is
\begin{equation}
N_{\rm pre}(t)= N(M_{\rm i},t_{\rm i})\,\delta
M_{\rm i}\,\exp\left\{-\int_t^{t_{\rm i}} \rf [M(t'),t')]\, \der
t'\right\}.\label{pre}
\end{equation}
Consequently, the PDF of formation times for halos with masses between
$M_{\rm i}$ and $M_{\rm i}+\delta M_{\rm i}$ at $t_{\rm i}$ is given by
\begin{eqnarray}
\Phi_{\rm f}(t|M_{\rm i})&\equiv&{1\over N(M_{\rm i},t_{\rm i})\,
\delta M_{\rm i}}\,{\der N_{\rm pre} \over\der t} =\nonumber \\
&=&\rf[M(t),t]\,\exp\biggl\{-\int_t^{t_{\rm i}} \rf[M(t'),t']\,\der
t'\biggr\}\,.
\label{dft}
\end{eqnarray}
The median for this distribution defines the typical halo formation
time.

The validity of all preceding analytical expressions was checked
against numerical simulations in RGS. The fact that the
simulations performed by \citet{LC94} used in that comparison have low
resolution by today standards should not affect the conclusions drawn
in RGS as they refer to the way halos grow and not to their inner
structure. However, it is worth re-examining, now with the help of
high-resolution simulations, the correct behavior of the mass
accretion rate (eq. [\ref{accr}]) determining the density profile of
halos in the present model (see next section). 

In Figure~\ref{1}, we compare the accretion histories predicted from
the MPS model with those obtained by W02 from the high-resolution
simulations of B01. The accretion histories of W02 have been converted
to the definition of halo mass used in the present work (see
\S~\ref{acc} for the values of $f$ and $\rou$ adopted). Note also that
since the latter accretion histories include both major and minor
aggregations they must be compared with our theoretical predictions
drawn from equation (\ref{accr}) {\it for $\delm$ equal to unity}. As
can be see, there is good agreement between theory and
simulations. Only for halo masses above the critical mass for
collapse, $M_\ast$, there is an increasing departure. This seems to
reflect the difficulty of the PS formalism to correctly model the mass
growth of halos towards large masses (see, e.g., Monaco 1997),
although no such departure is detected in Fig. 3 of RGS.

\section{THE DENSITY PROFILE SET BY ACCRETION}\label{acc}

The rearrangement, through violent relaxation, of a halo at its
formation produces a more or less spherical system with a new density
profile independent of the past history of the halo. On the other hand,
accretion consists of frequent multiple captures of very small
halos. Thus, the graininess of accreted matter can be neglected in a
first approximation. The resulting configuration (i.e., a central
relaxed spherical object surrounded by a rather smooth distribution of
matter falling into it) approximately satisfies the conditions met in
secondary infall. Consequently, halos grow inside-out during accretion,
making it possible to determine the growth of their density profile
from the accretion rate derived in the preceding section.

\subsection{Predictions at $z=0$}

Consider a halo with current mass $M$ having formed at $\ti$. During
the subsequent accretion phase, the halo follows the $M(t)$ track
solution of the differential equation (\ref{act}). According to the
inside-out growth condition, the mass accreted at any moment $t$ is
deposited at the instantaneous radius $R(t)$ of the system without
altering the inner density profile. Consequently, we have
\begin{equation}
M(t)-\Mi=\int_{\Ri}^{R(t)} 4 \pi r^2 \rho(r)\,\der r\label{mt}\,,
\end{equation} 
with $\Mi$ and $\Ri$ the mass and radius at formation and $\rho(r)$
the density profile developed since that moment. By differentiating
equation (\ref{mt}) and taking into account the definition of the
virial radius
\begin{equation}
R(t)=\left[ {3\,M(t) \over f 4 \pi \rou(t)} \right]^{1/3} 
\label{rad}
\end{equation} 
and equation (\ref{act}), we are led to the expression
\begin{equation}
\rho(t)= f \rou(t) \left\{ 1-{M(t) \over \ra[M(t),t]}
{\der \ln \rou \over \der t} \right\}^{-1}    
\label{dens}
\end{equation}
giving the density at $r=R(t)$. Equations (\ref{rad}) and (\ref{dens})
therefore define, in the parametric form, the density profile
developing by accretion from the initial seed at $\ti$, at any radius
$r\geq\Ri$. Notice that, owing to the abovementioned dispersion of
real $M(t)$ tracks around the solution of equation (\ref{act}), the
previous theoretical profile represents, like the NFW profile, the
{\it average\/} density profile of halos with $M$ at $t$, with
individual profiles fluctuating around it.

The shape of this theoretical profile depends on the values of $f$ and
$\rou$ entering the definition of the virial radius
(eq. [\ref{rad}]). The need to choose appropriate values for these two
quantities in order to obtain good predictions from the PS model is
not new. For instance, the conditional as well as the unconditioned
mass functions of halos showing a given abundance depend on the radius
adopted to define the halo mass. In Figure~\ref{2}, we plot the
theoretical density profiles predicted from $\rou=\rocr$ and
$\bar\rho$ in a flat, $\Omega_{\rm m}=0.25$ CDM cosmology assuming
that accretion operates from a very small seed. As can be seen,
$\rou=\bar\rho$ leads to a monotonous decreasing density profile,
while $\rou=\rocr$ leads to a density profile which tends to level off
at those radii corresponding to cosmic times when $\Lambda$ becomes
dynamically dominant. Since such radial behavior is not observed in
real or simulated halos, we conclude that the best value of $\rou$ to
use is $\bar\rho$. The resulting profile can then be further
truncated, of course, at any smaller radius if necessary. This will be
done when comparing our theoretical profile with the NFW profile that
uses $\rou=\rocr$. Concerning $f$, the theoretical density profile is
little sensitive to it, any usual value yielding similar
results. Accordingly, unless we state otherwise, we will use $f=200$
as in NFW.

The theoretical density profile also depends (through
eq. [\ref{accr}]) on the value of $\delm$, the effective threshold for
mergers. As explained in RGS, the comparison between theory and
simulations in terms of halo masses alone puts no constraint on this
parameter. In contrast, the inner structure of halos can be used to
fix the value of this parameter. In Figure~\ref{3}, we show the effects of
varying $\delm$ within the range [0.3,0.7] around the best value of
$0.5$ (see below), which we use hereafter.

Once the values of $\delm$, $\rou$, and $f$ have been fixed, the
theoretical density profile of halos with $M$ at $t$ having evolved by
accretion from the initial seed formed at the last major merger can be
inferred from equations (\ref{rad}) and (\ref{dens}). In
Figure~\ref{4}, we show the profiles obtained assuming {\it pure
accretion}, meaning that the last major merger leading to the formation
of the halo took place long ago, and we compare them to the
corresponding NFW profiles. We plot only three cosmologies, although
similar results are obtained in any of the cosmologies studied by
NFW. Taken as a whole, there is a good agreement between the
theoretical and empirical profiles down to $r\approx 10^{-2}R$ for
halos spanning at least four decades in mass. However, a systematic
tendency is observed for the accordance to worsen beyond $M_\ast$,
becoming substantial at about 10 times $M_\ast$. This departure is
likely caused by the abovementioned deviation for large halo masses
between the accretion histories predicted by the MPS model and those
obtained in high-resolution simulations. But the most massive halos are
scarce and seldom virialized (typically they have formed recently), so
the disagreement between our model and the data is not important in
practice.

Both the theoretical and empirical profiles shown in Figure~\ref{4}
for different halo masses are completely fixed in shape {\it as well
as in scale}. This means that our approach yields not only a density
profile similar to the NFW one (at least for moderate and small halo
masses), but also good values of $c$. In this respect, it is
worth noting that the NFW expression is a fit, i.e., only a good
approximation to the real density profile of simulated halos, and that
the empirical mass-concentration relation is inferred from a small
number of halos of different masses in each cosmology (the few points
in the mass-concentration diagram in Fig.~\ref{5}). This affects, of
course, the comparison between the theoretical and empirical halo
profiles performed in Figure~\ref{4}. Hence, to better assess the
quality of the predictions of our formalism it is preferable to
adjust, through $\chi^2$ minimization, the theoretical profiles
corresponding to different halo masses by the NFW expression, as if
they were the profiles of individual simulated halos, and compare the
resulting theoretical mass-concentration relation with \nbody data.

As shown in Figure~\ref{5}, the theoretical mass-concentration
relations above $\sim 10~M_\ast$ deviate strongly from the empirical
data, particularly for power-law spectra. As indicated by the $\chi^2$
values, this departure is due to the progressive disagreement between
the theoretical profile and the NFW expression with increasing halo
mass. In contrast, below $\sim 10~M_\ast$, the empirical
mass-concentration relationships for the cosmologies studied by NFW are
well bracketed for $\delm$ between 0.3 and 0.7, with the lower and
upper bounds tending to yield too large and too small concentrations,
respectively. Hence, were the effective frontier between accretion and
mergers lowered to exclude intermediate captures, the predicted
concentration would be larger than that found in simulated halos, in
agreement with the results of \citet{ns99}, \citet{kull00} and
\citet{pgrs}. In \S\ \ref{merg} we will provide, however,
justification for the classification of intermediate captures as
accretion.

\subsection{Other redshifts}\label{cz}

Since our model can be applied to any redshift, we can also derive the
redshift dependence of $c$ for halos of a fixed mass and compare it to
that found by B01 in \nbody simulations of the $\Lambda$CDM
cosmology. These authors adopted a halo radius $R$ defined according to
equation (\ref{rad}) with $\rou=\bar\rho$ as here, but $f$ varying from
337 to 178 as $z$ shifts from 0 to 5 according to the top-hat collapse
model in the $\Lambda$CDM cosmology. Therefore, in order to carry out
this comparison one can derive the theoretical profile using $f=200$ as
above, and then truncate or extend it to the same $z$-dependent radius
as in B01 or, alternatively, one can take advantage of the fact that
the theoretical profile is little sensitive to $f$ and derive it
directly using the same $f(z)$ of B01. In the former case, the halo
mass can only be known a posteriori after the appropriate truncation or
extension of the halo density profile, while, in the latter case, it
coincides with the mass used by B01. For this reason, we have chosen
to follow the latter approach.

In Figure~\ref{6} we show the predicted $c(z)$ relation and compare it
to that found by B01 for a halo mass equal to $5.55\times 10^{12}$
M$_\odot$, the central logarithmic value of the mass range $2.14\times
10^{11}$ -- $1.43\times 10^{14}$ $M_\odot$ studied by these
authors. This mass is small enough, along the whole range of redshifts
involved in the comparison, for the theoretical profile to always be
well adjusted by the NFW expression while, at the same time, it is
sufficiently large to guarantee that simulated halos have a large
number of particles and, hence, that their concentration is well
determined. The agreement between the predicted and empirical $c(z)$
relations is good. We want to stress that this is the first correct
prediction of $c(z)$ made by a physical model. Previous predictions
based on the proportionality $\roc\propto \rou(\ti)$ proposed by NFW
(corresponding to a universal dimensionless density profile of halos at
formation; SSM) have been proved, on the contrary, not correct
regardless of the exact definition of $\ti$ used (see Fig.~\ref{6}),
while other analytical expressions which correctly fit the empirical
function $c(z)$ are but toy models (B01; \citealt{ens01}).

What is wrong then with the assumption $\roc\propto \rou(\ti)$?
According to the present model, all halos lying, at different epochs,
along a given accretion track $M(t)$ have the same density profile,
though truncated at their respective virial radii. The fact that the
theoretical density profile of halos at $z=0$ is reasonably well
fitted by the NFW expression over the whole radial range implies that
all those halos have essentially the same values of the scale
parameters $\roc$ and $\rs$ (though not of $c$, which varies with $z$
owing to the time dependence of $R$). In other words, $\roc$ and $\rs$
do not change along accretion tracks. Consequently, they remain
constant for halos evolving by accretion (in agreement with the
results of \nbody simulations; \citealt{zhao03}). In particular, the
value of $\roc$ {\it does not depend on the formation time of
halos\/}, but on their (last) accretion track.

\section{THE EFFECTS OF MERGERS AND INTERMEDIATE CAPTURES}\label{merg}

The preceding theoretical profile has been derived assuming pure
accretion and the inside-out growth of halos during such a
process. This presumes: 1) that halos have not undergone any merger
for a very long period of time and 2) that the capture of clumps with
intermediate masses has a negligible contribution in the accretion
process. But both conditions are far from realistic in hierarchical
cosmologies. For instance, \nbody simulations \citep{T97} show that
the average numbers of aggregations producing, in the standard CDM
cosmology, a fractional mass increase in halos of cluster scale above
0.1, 0.2, and 0.5 along the age of the universe are, respectively,
$13.7\pm1.4$, $7.4\pm1.0$, and $2.3\pm0.5$. These values reduce to
$2.7\pm0.5$, $1.5\pm0.4$, or $0.1\pm0.1$ if we only consider the time
elapsed since the formation of the halos, defined as in the original
extended PS model, i.e., the time that half the mass of the halo was
first collected in some progenitor \citep{LC93}, while our definition
of halo formation time (see \S~\ref{mps}) leads to intermediate
values.

\subsection{Mergers}

Though rare, mergers are actually frequent enough to invalidate the
assumption of pure accretion. This is illustrated in
Figure~\ref{4}. Arrows mark the radii of the seeds formed at the
typical last merger of halos inside which the density profile predicted
assuming pure accretion should be replaced by that yielded by violent
relaxation. However, as mentioned in \S~\ref{intro}, the latter profile
is eventually very similar to the one the system would have developed
through pure spherical infall. We want to emphasize that this is not a
consequence of our modeling but an empirical fact drawn from
high-resolution cosmological simulations. As shown next, such a
``coincidence'' emanates from the fact that violent relaxation takes
some finite time to proceed during which accretion keeps going on. The
final density profile then adopts the {\it unique} inner structure
compatible with the boundary conditions imposed by the accreting
layers.

We begin by showing first that the velocity dispersion profile
$\Sigma(r)$ of halos in the outer accreted region is uniquely
determined. It could be derived, of course, from the Jeans equation
(\ref{heq}) provided we knew the value it takes at some given
radius. But this is information that we do not have a priori (we are
avoiding any unjustified assumption at infinity). What we only know is
that the velocity dispersion profile of seeds emerging from violent
relaxation in real non-isolated systems deviates from the uniform one
expected for ideal isolated systems owing to the boundary conditions
imposed by the infalling matter. This suggests the following
iterative way to determine the value of $\Sigma$ at $r=\Ri$, the
frontier between the inner seed and the outer accreted region.

The zero-order solution of $\Sigma^2$ at $r=\Ri$ is given by the
squared value of the uniform velocity dispersion that violent
relaxation tends to establish, given by equation (\ref{heq}) with null
logarithmic derivative of $\Sigma^2$ and logarithmic derivative of
$\rho$ corresponding to the known outer density profile. But violent
relaxation takes a finite time to proceed, during which the
instantaneous radius of the halo shifts outwards. Consequently, the
same condition will hold at the new edge of the system after some
arbitrarily small time. We can therefore derive in the same way the
zero-order value of $\Sigma^2$ at this new edge of the system. Those
two values of $\Sigma^2$ can then be used to determine the zero-order
logarithmic derivative of $\Sigma^2$ at $r=\Ri$ and, by substituting
this derivative into the equation (\ref{heq}), to infer the
first-order value of $\Sigma^2$ at that point. This iterative
procedure rapidly converges to the wanted value of $\Sigma^2(\Ri)$.

Once we know the value of $\Sigma(\Ri)$, we can infer the velocity
dispersion profile in the outer accreted region where the density
profile is known (eqs.~[\ref{rad}] and [\ref{dens}]) from the Jeans
equation (eq.~[2]). But this velocity dispersion profile must coincide
with the one that can be obtained by directly applying the previous
iterative procedure at any point of the outer region as both solutions
satisfy, by construction, the Jeans equation for the same density
profile and have identical values at $\Ri$. To sum up, thanks to the
uniform velocity dispersion profile that violent relaxation {\it
tends\/} to establish in the inner seed, the velocity dispersion
profile emerging in the outer accreted region is fully determined,
turning out to be also {\it as close to uniform as allowed by the
density run there}. In Figure~\ref{7}, we compare, for the same halo
masses and hierarchical cosmologies as in Figure~\ref{4}, the velocity
dispersion profile of simulated halos \citep{CL97} with the
theoretical one inferred following the previous iterative procedure
over the whole radial range as would correspond to the case of pure
accretion.  As can be seen, the agreement between the theoretical and
empirical velocity dispersion profiles is as good as in the case of
the density profiles.

We can now proceed with our reasoning and focus on the density profile
of the inner seed. Since violent relaxation lasts for some time during
which matter is continuously falling into the halo, the final mass
distribution adapts to the boundary conditions imposed by this matter
infall. Thus, it must be in steady state, compatible with the shell
crossing of infalling layers, and with a velocity dispersion as close
to uniform as possible. But these are the same conditions satisfied by
the halo during the accretion process. It is therefore well understood
why the density profile emerging from a major merger coincides with
the profile that would have developed by means of pure
accretion. Furthermore, since the concept of orbital decay is
meaningless during violent relaxation, there is no difference between
intermediate and small captures during that process apart from the
distinct mass increase they produce. Therefore, intermediate captures
must be included in the formal accretion rate leading to the same
density profile as violent relaxation.

It might be argued that the previous reasoning presumes that violent
relaxation proceeds to completion, while the relaxation time, equal
to, say, three crossing times of the system at the time of the major
merger,
\begin{equation}
t\rel(\ti)\approx 3\,\frac{2\pi\, R(\ti)}{\Vc(\ti)}\approx 0.66
\left[G\bar\rho(\ti)\right]^{-1/2}\,,\label{trel}
\end{equation}
with $\Vc=(GM/R)^{1/2}$ the circular velocity of the halo, is long
enough for that condition not always to be satisfied. However, the NFW
expression describes the average profile of halos having a relaxed
appearance. That is, those halos observed before violent relaxation
has gone to completion are not taken into account.

\subsection{Intermediate Captures}

Provided that halos do not undergo intermediate captures after the
completion of violent relaxation at their last major merger, their
density profile will develop inside-out as explained in \S\
\ref{acc}. Strictly, there is no need, in the present case, for
accretion to include intermediate captures, although given that the
outermost profile is quite insensitive to the value of $\delm$ used
(see Fig.~\ref{3}) this makes almost no difference in practice. If, on
the contrary, halos undergo intermediate captures in such a late phase,
as one would naively expect from the larger frequency of intermediate
captures as compared to major ones, the density profile will become
cuspier and deviate from the NFW profile.

What is therefore crucial to understand the shape of the NFW profile
of relaxed halos is to see that such intermediate captures are quite
improbable. This is not in contradistinction with the relatively high
average numbers quoted by \citet{T97}. Here we must only consider
those intermediate captures restricted to occur after the completion
of the violent relaxation of the halo accompanying its last merger
{\it and\/} early enough for the captured clump to have completed its
orbital decay by the time the halo is observed. This latter bound is
necessary, indeed, to guarantee that the halo does not show any
substructure and can then be identified as a relaxed system. The low
frequency of intermediate captures subject to these two constraints
seems to be supported by the results of \nbody simulations
\citep{aygm}.

Let us therefore calculate the probability ${\cal P}(M)$ that the last
intermediate merger of a halo with $M$ at $t$ occurred after the
completion of violent relaxation and the time at which the eventual
intermediate merger should take place for the merged clump to reach
the halo center by the time the halo is observed. This probability is
given by
\begin{eqnarray}
{\cal P}(M)&=&\int_0^{t_0} \Phi(\ti|M)\,\der\ti 
\int_{\Delta\minb(\ti)}^{\delm/(1+\delm)}\der \Delc \nonumber \\ 
&\times& \int_{\ti+t\rel(\ti)}^{t_0-t_{\rm
fric0}(\Delc)} P\last(\Delc,t|M)\,\der t\,,\label{paff}
\end{eqnarray}
with the integrand giving the joint probability that the halo was
formed from $\ti$ to $\ti+\der\ti$ and underwent the last intermediate
merger with a clump of instantaneous fractional mass between $\Delc$
and $\Delc+\der\Delc$ in the infinitesimal interval of time around
$t$. This joint probability is simply equal to the product of the
probabilities $\Phi(\ti|M)\,\der\ti$, given in equation (\ref{dft}),
and $P\last(\Delc,t|M)\,\der\Delc\,\der t$, calculated in the
Appendix. 

In equation (\ref{paff}), $\ti+t\rel(\ti)$ is the time of completion
of violent relaxation after the last major merger at $\ti$, with the
relaxation time $t\rel$ given by equation (\ref{trel}), and $t_0-t_{\rm
fric0}(\Delc)$ is the time at which the intermediate merger should
occur for the clump to have reached the halo center at $t_0$. The
characteristic time of orbital decay of a clump of mass $M\sat$ in a
circular orbit of radius $r$ is (e.g., \citealt{binn87})
\begin{equation}
t\orb(r)\equiv \frac{r}{\dot
r}=\frac{1}{0.43\,\ln[M(r)/M\sat]}\frac{\Vc(r) \,r^2}
{G\,M\sat}\label{tod}\,,
\end{equation}
with $\Vc(r)=[GM(r)/r]^{1/2}$ the circular velocity at $r$. Hence, by
integrating $t\orb\,\der r$ under the approximation of a singular
isothermal halo (for which the integral is analytical) from 0 to the
radius of the initial orbit, $R(t)$, at the time $t$ of the merger, we
have an estimate of the time spent by the clump to spiral to the halo
center owing to dynamical friction
\begin{eqnarray}
t\fric(t\merg,\Delc)&=&\frac{1}{0.43\,\ln[M(t)/M\sat]}\frac{\Vc(t\merg)\,
R(t\merg)^2} {2\,G\,M\sat} =\nonumber \\
&=&-\frac{0.13\Delc^{-1}}{\ln(0.3\Delc)}
\left[G\bar\rho(t\merg)\right]^{-1/2}\label{df}\,,
\end{eqnarray}
where $M(t)$ and $\Vc(t)$ are, respectively, the total mass and
circular velocity of the halo at the time of the merger. Thus, $t_{\rm
fric0}(\Delc)$ is given by the previous expression for a time $t$
solution of the implicit equation $t+t\fric(t,\Delc)=t_0$. Note that
in deriving equation (\ref{df}), we have assumed, for simplicity, that
$M\sat$ remains constant and equal to about 30\% of the mass of the
merged clump after the initial tidal stripping, i.e.,
$M\sat=0.3\,M(t\merg)\,\Delc$, with $\Delc$ the fractional mass
increase of the halo relative to the final object. According to
\citet{DDH03}, the mass of clumps is reduced to 30\% of its initial
value after their orbits have already achieved a substantial
decay. But these authors assume clumps with a density profile steeper
than the NFW one at large radii, which means that their clumps are
more difficult to tidally strip than real ones. On the other hand, we
expect the real orbits of merged clumps to be typically elliptical
rather than circular as we are assuming here, which should diminish
their time of orbital decay.  However, this effect should be balanced
by the fact that, in the elliptical case, clumps fall deeper in the
halo at their pericenter and, hence, are more severely truncated by
tides since the beginning.

The upper bound for $\Delc$ in equation (\ref{paff}) is not
$\delm=0.5$ but $\delm/(1+\delm)=0.33$ since the fractional mass
increase $\Delc$ is relative to the mass $M(t)$ of the halo {\it
after\/} the merger. The lower bound $\Delta\minb$ is the value of
$\Delc$ yielding a characteristic time of orbital decay for the
initial orbit (eq.  [\ref{tod}] with $r=R[t])$, equal to
$t_0-[\ti+t\rel(\ti)]$, that is, the solution of the implicit equation
\begin{equation}
-\frac{0.27\Delta\minb^{-1}}{\ln(0.3\Delta\minb)}
\frac{1}{\sqrt{G\bar\rho\{t_0-[\ti+t\rel(\ti)]\}}}=
t_0-[\ti+t\rel(\ti)].
\end{equation}
Values of $\Delc$ smaller than this limit correspond to clumps that
do not suffer any significant orbital decay.

In Figure~\ref{8}, we show, for different halo masses and cosmologies,
the quantity ${\cal P}$ giving the probability that intermediate
captures can affect the halo density profile that would emerge from the
last major merger and the subsequent accretion phase, calculated in
\S\ \ref{acc}. As can be seen, this probability decreases with
increasing mass. It becomes negligible for halos more massive than
about $M_\ast$ in all cosmologies, but even for the smallest halo
masses considered by NFW, equal to $\sim 10^{-2}M_\ast$ for CDM
spectra or $\sim 10^{-1.5}M_\ast$ for power-law ones, it is quite
small. Certainly, the exact values shown in this plot should not be
taken too literally given that they depend on the exact definitions of
$t\rel$ and ${t\fric}$ adopted. But it is equally true that equation
(\ref{paff}) gives an upper bound for ${\cal P}$ since, in its
derivation, it has been implicitly assumed that the mass increase in
captures comes from one unique clump, while it might also be produced
by two or more less massive clumps not falling then into the category
of intermediate captures. What is more important, more than one
intermediate merger is required to affect appreciably the density
profile, while the probability that two or more intermediate captures
take place in the right time interval is obviously much smaller.

\section{SUMMARY AND DISCUSSION}\label{diss}

We have presented a simple analytical model for the density profile
developed by halos during accretion based on their inside-out growth in
that regime. The resulting density profile is found to be similar to
the NFW profile in agreement with previous works. However, contrarily
to these same works, we also find good agreement between theory and
simulations concerning the mass-concentration relations of halos at
$z=0$. More importantly, our physical model correctly predicts the
mass-concentration relation at any redshift. An analytical model such
as the one presented here, able to make reasonable predictions on the
density profile of halos at any epoch and in any cosmology, should be
then very useful in the modeling of galaxy formation and
evolution. Caution must be taken, however, in using the present model
on halos more massive than $\sim 10 M_\ast(t)$, with $M_\ast(t)$ the
typical mass for collapse at $t$. For masses this large, the
theoretical profile shows a substantial departure from the empirical
one likely due to a deficient modeling of halo growth by the PS
formalism at the large mass end.

The success of our model compared to previous ones appears to rely on
the inclusion of intermediate-mass captures in the accretion regime
despite the fact that they do not satisfy the inside-out growth
condition. This is justified by the realization that the halo density
profile is actually not set during accretion (except for the narrow
outermost radial range, in logarithmic units), but at the time of the
last major merger, both processes leading to halos with essentially the
same shape. Given that there is no difference between the dynamical
effects of minor and intermediate captures during the violent
relaxation resulting from mergers, it is then understandable why
intermediate captures must be included formally in the accretion
process.

The fact that major mergers yield essentially the same density profile
than pure accretion is an important result of high-resolution numerical
simulations that has been pointed out only by a few authors (W02;
\citealt{Moore99,HJS99b}). We have shown here that this unexpected
coincidence is simply due to the unicity of the steady structure
compatible with the boundary conditions imposed by the ever falling
surrounding material. Thus, we have been able to reconcile the
well-known result that pure accretion (or spherical infall) yields
density profiles of the NFW form with the fact that, in hierarchical
cosmologies, halos endure from time to time important mergers (or short
periods of intense accretion) that bring them far from equilibrium and
cause the rearrangement of their preceding structure. Besides, we have
demonstrated that there is no time for intermediate captures to alter
the density profile \`a la NFW established after the last merger
without giving the halo a non-relaxed appearance.

Finally, we want to remark that our aim here was not to infer an
accurate density profile for dark-matter halos, but to investigate its
possible origin. For this reason we have considered the simplified case
of pure dark-matter halos while, in the real universe, about $10-15\%$
of the halo mass is in the dissipative baryonic component. This might
have appreciable effects on the density profile of real
systems. Likewise, we have assumed spherical symmetry and neglected
halo rotation, as well as the possible anisotropy of the local velocity
tensor. Real halos have, on the contrary, some angular momentum and are
immersed in large filamentary structures making them accrete matter
preferentially along one privileged direction \citep{West94} and feel
the tidal field of such anisotropic structures \citep{ss93}. Moreover,
even with perfect spherical symmetry, some velocity anisotropy will
prevail in the halo outskirts, as observed, owing to the distinct
evolution of the radial and tangential velocity dispersions of
infalling layers. We note, however, that our density profile is
independent of the actual degree of anisotropy of the velocity tensor.



\acknowledgments

We thank Guillermo Gonz\'alez-Casado for helpful discussions and Julio
Navarro for kindly providing the data of the mass-concentration
relations. We also thank Julio Navarro, James Bullock, and their
respective collaborators for making public the codes necessary to
reproduce their results. This work was supported by the Spanish DGES
grant AYA2000-0951. A.M. acknowledges the hospitality of the CIDA staff in
M\'erida (Venezuela) where part of this work was carried out.

\appendix

\section{PROBABILITY OF THE LAST INTERMEDIATE MERGER}

Let us consider the rate of intermediate captures up to some given
fractional mass increase $\Delc$ of halos with final mass $M'$
at $t$
\begin{equation} 
\ri(M',\le\Delc,t)= \int^{M'\Delc}_{M'\Delc\minb} r\PS
(M,M',t)\,\frac{N(M,t)}{N(M',t)}\,\der M\,.\label{form2}
\end{equation}

Following the same reasoning leading to the PDF of formation times, but
adapted to the previous range of intermediate captures, we have that
the cumulative number density of current halos with masses in the
arbitrarily small range $M$ to $M+\delta M$ that have undergone the
last intermediate merger up the fractional mass increase $\Delc$ at
some time smaller than $t$ is
\begin{equation}
N_{\rm last}(\le\Delc,\le t\,|\,M)= N(M,t_0)\,\delta M
\,\exp\left\{-\int_t^{t_0} \ri[\tilde M(t'),\le\Delc,t')]\, \der
t'\right\},\label{last}
\end{equation}
where $\tilde M(t)$ stands for the accretion track calculated using a
modified threshold for mergers $\tilde\delm$ corresponding to
$\Delc\minb$ to take into account that, on the final phase, accretion
should not include intermediate captures. Hence, the PDF of times that
halos with current mass $M$ underwent their last intermediate merger
yielding a fractional mass increase from $\Delc$ to $\Delc+\der\Delc$
is simply
\begin{equation}
P\last (\Delc,t|M)\,\der\Delc\equiv{1\over N(M,t_0)\,\delta
M}\,\left|\frac{\partial^2}{\partial t\,\partial
\Delc} N\last(\le\Delc,\le t|M)\right|\,\der\Delc\,. 
\label{dft2}\nonumber
\end{equation}

\clearpage

\renewcommand{\thefigure}{\arabic{figure}}

\begin{figure}
\centerline{
\epsfxsize=16cm
\epsffile{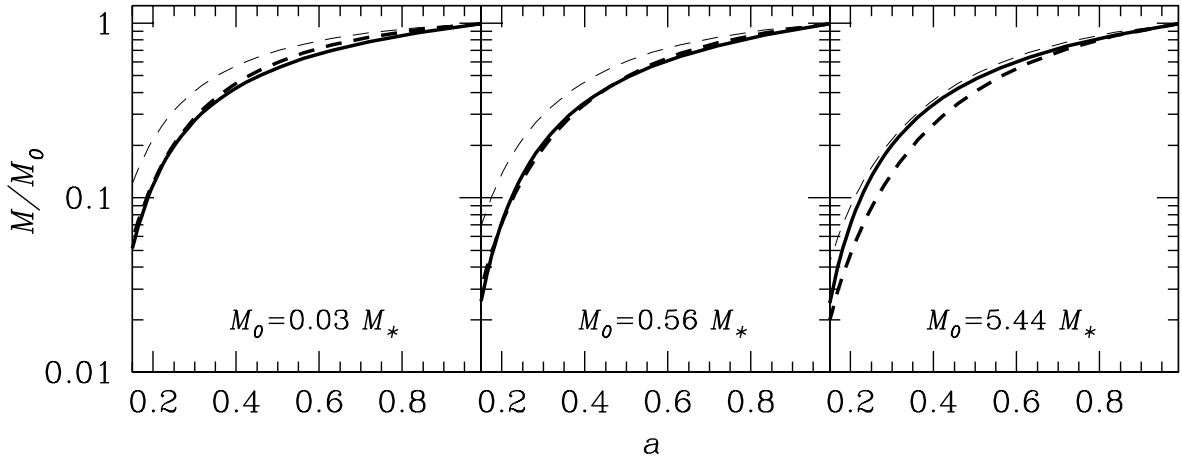}}
\vskip -9.5cm
\caption{Mass accretion histories for three halos in the flat
$\Lambda$CDM cosmology studied by B01. The MPS predictions for $\delm$
equal 1 (thick dashed lines) and 0.5 (thin dashed line) are compared
to the analytical expression proposed by W02 to fit the mass accretion
histories drawn from the \nbody simulations of B01 (full
line). $M_\ast=2.39\times 10^{13}$ M$_\odot$ is the typical mass for
collapse of density fluctuations at $z=0$ for this cosmology.}
\label{1}
\end{figure}

\begin{figure}
\centerline{
\epsfxsize=16cm
\epsffile{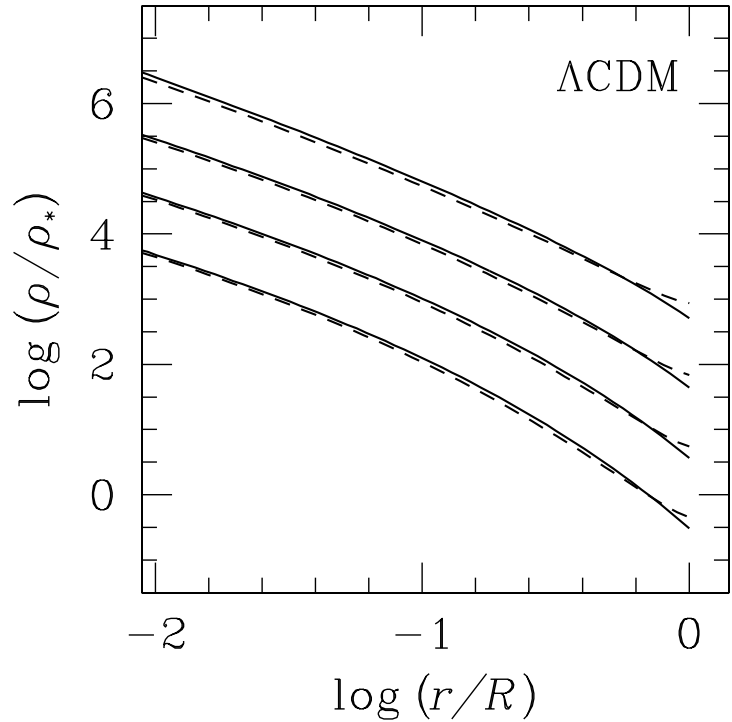}}
\vskip -3cm
\caption{Theoretical density profiles for halos at $z=0$ with masses
equal to 10$^{-2} M_\ast$, 10$^{-1} M_\ast$, $M_\ast$, and 10 $M_\ast$
(from bottom to top) in a flat, $\Omega_{\rm m}=0.25$, CDM model
inferred using $\rou=\rocr$ (dashed lines) and $\rou=\bar\rho$ (solid
lines). We are using the same cosmological parameters and delimitation of
halos as in NFW and $\delm=0.5$. $M_\ast$ is the typical mass for
collapse of density fluctuations and $\rho_\ast$ stands for $\rocr
M_\ast/M$.}
\label{2}
\end{figure}

\begin{figure}
\centerline{
\epsfxsize=16cm
\epsffile{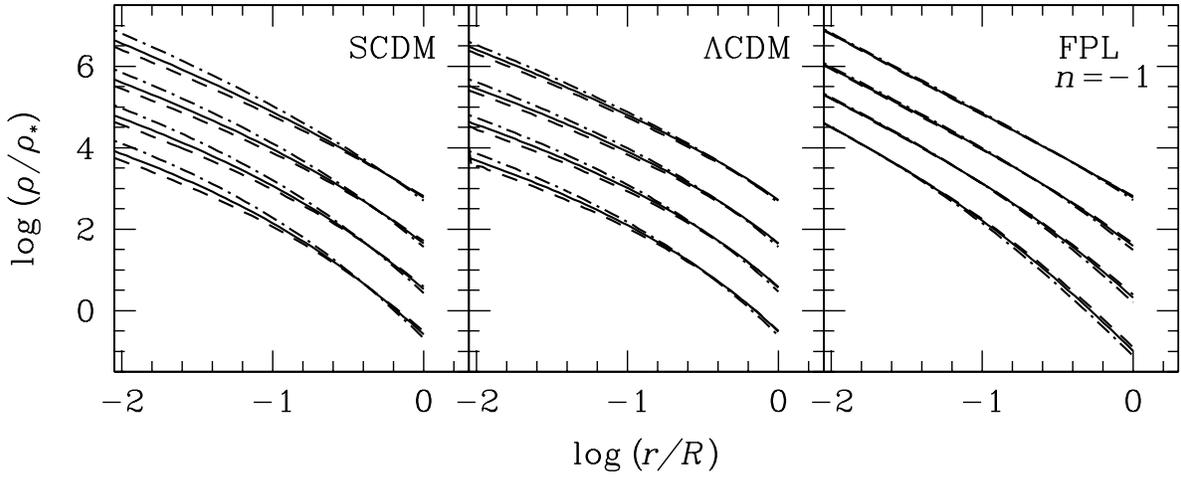}}
\vskip -9cm
\caption{Theoretical density profiles for halos at $z=0$ with the same
masses as Fig.~\ref{2} obtained for different values of the threshold
for mergers: $\delm=0.3$ (dot-dashed lines), $0.5$ (solid lines), and
$0.7$ (dashed lines). The three cosmologies represented are a flat,
$\Lambda=0$, CDM model (SCDM), a flat, $\Omega_{\rm m}=0.25$, CDM model
($\Lambda$CDM), and a flat, $\Lambda=0$, model (FPL) with power-law
spectrum of density fluctuations of index $n=-1$. All model parameters
are the same as in the corresponding cosmologies of NFW.}
\bigskip\label{3}
\end{figure}

\begin{figure}
\centerline{
\epsfxsize=16cm
\epsffile{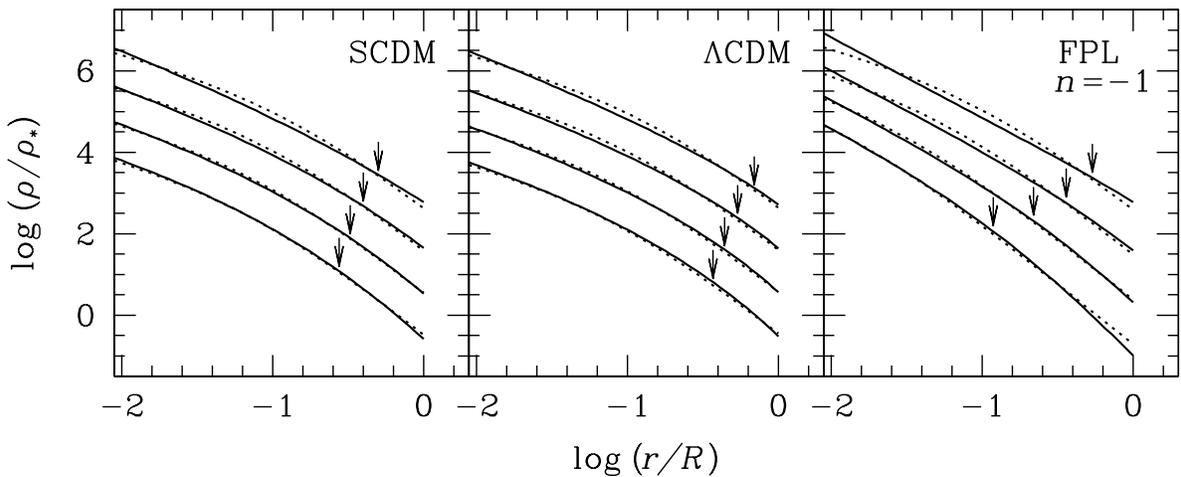}}
\vskip -9cm
\caption{Theoretical (solid lines) and NFW (dotted lines) density
profiles for halos at $z=0$ for the same masses used in the previous
Figures and the same cosmologies depicted in Fig.~\ref{3}. Arrows mark the
halo radii at the {\it typical\/} halo formation times.}\label{4}
\end{figure}

\begin{figure}
\centerline{
\epsfxsize=16cm
\epsffile{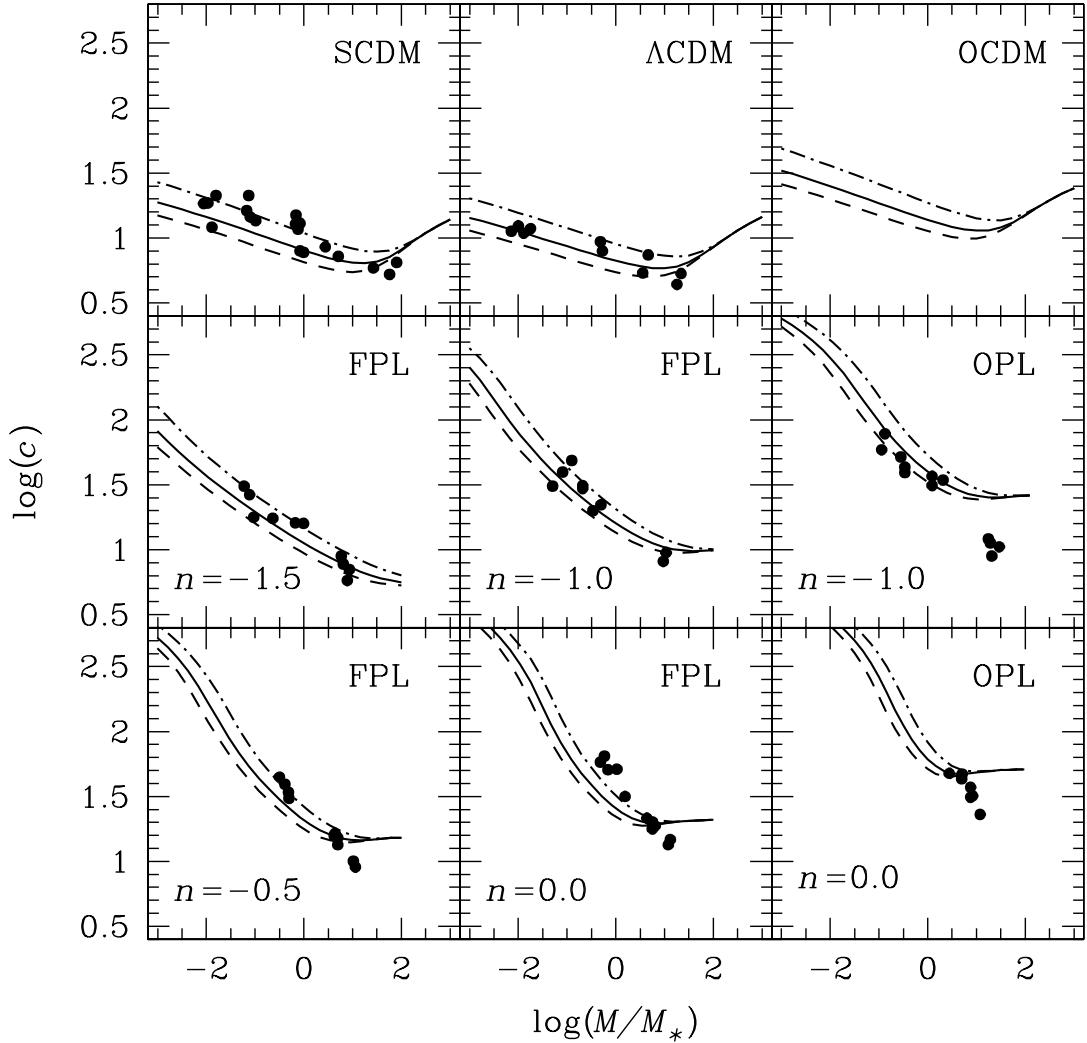}}
\vskip -1cm
\caption{Mass-concentration relation for halos at $z=0$ obtained by
fitting the theoretical density profiles derived in the present paper
by the NFW analytical expression. The solid line corresponds to
$\delm=0.5$, while the dot-dashed and dashed lines correspond to the
rather extreme values of 0.3 and 0.7, respectively. The dark points
are those obtained by NFW from high resolution numerical
simulations. We plot all the cosmologies studied by these authors,
which, apart from those mentioned in the previous Figures, are two
$\Omega_{\rm m}=0.10$ power-law models (OPL) with $n=0,\ -1$, and three
more flat, $\Lambda=0$, power-law models (FPL) with $n=0,\ -0.5,\
-1.5$. We also show the predictions for an open, $\Omega_{\rm
m}=0.25$, CDM cosmology (OCDM), normalized as the $\Lambda$CDM model,
not included in the study by NFW.}\label{5}
\end{figure}

\begin{figure}
\centerline{
\epsfxsize=16cm
\epsffile{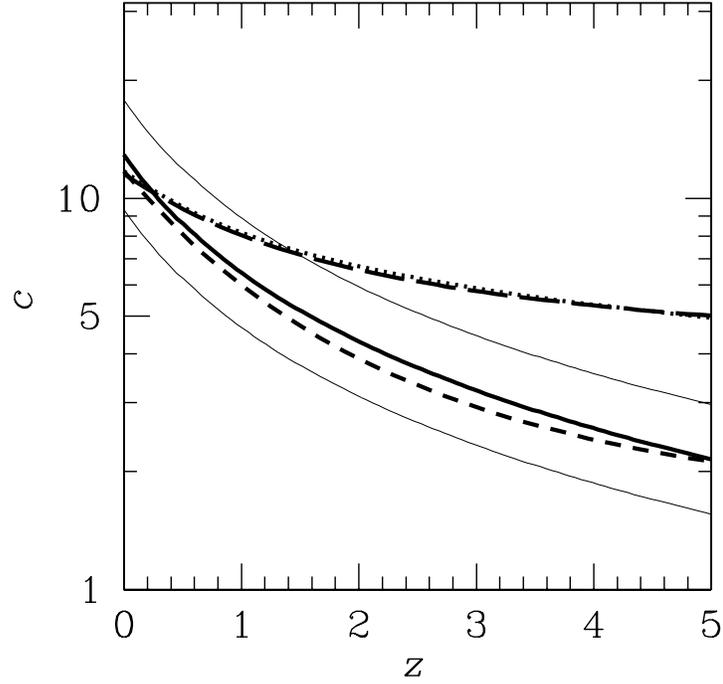}}
\vskip -3.3cm
\caption{Concentration vs. redshift in the $\Lambda$CDM cosmology
studied by B01 for a fixed halo mass $M$ equal to
$5.55\times 10^{12}$ $M_\odot$. The function proposed by
B01 to fit their empirical data (solid line) and the
corresponding intrinsic 68\% spread (thin solid lines) are compared to
our theoretical prediction (dashed line) and the $c(z)$ dependence
suggested by NFW (long-dashed line) and SSM (dotted line).}\label{6}
\end{figure}

\begin{figure}
\centerline{
\epsfxsize=16cm
\epsffile{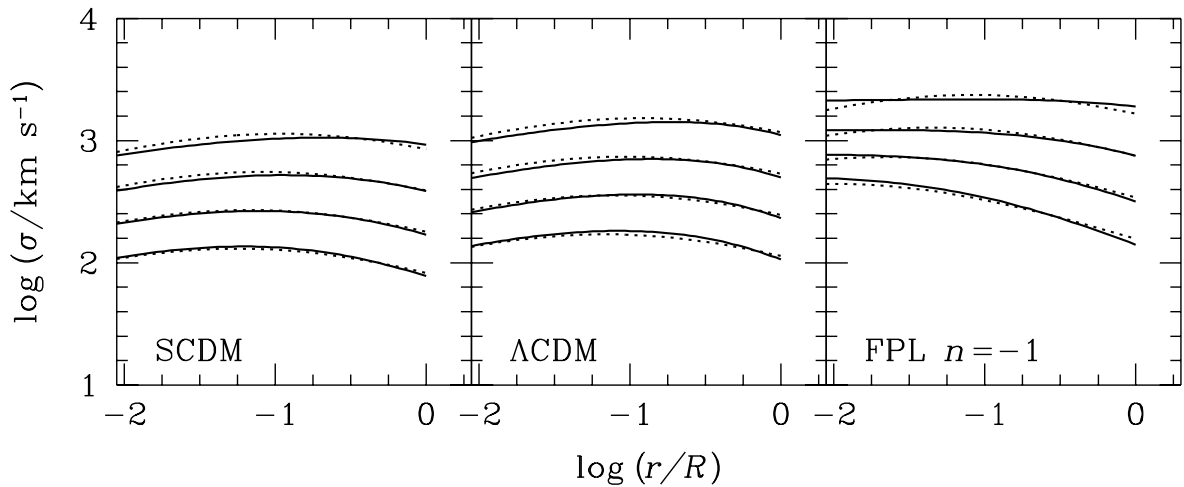}}
\vskip -9cm
\caption{Same as Fig.~\ref{4} but for the velocity dispersion
profile.}\label{7}
\end{figure}

\begin{figure}
\vskip -2cm
\centerline{
\epsfxsize=16cm
\epsffile{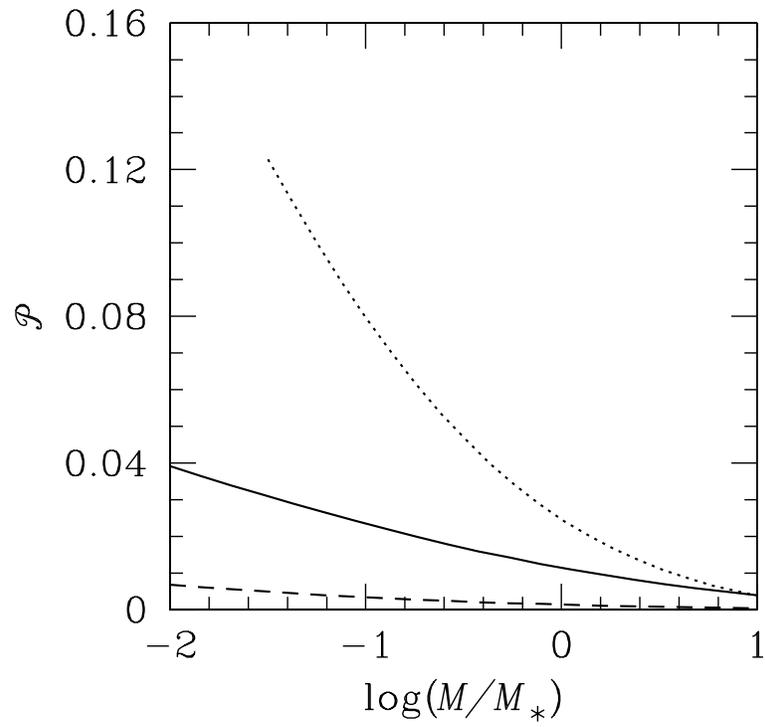}}
\vskip -3cm
\caption{Probability that currently relaxed halos with mass $M$ have
undergone some merger with a clump of mass in the range required to
modify the NFW-like profile arising from their last major merger and
the subsequent accretion phase. The different curves correspond to the
cosmologies plotted in Figs. 3, 4, and 7: SCDM (solid line),
$\Lambda$CDM (dashed line), and FPL $n=-1$ (dotted line).}\label{8}
\end{figure}


\end{document}